\documentclass{article}

\usepackage[final]{nips_2016}

\usepackage[utf8]{inputenc} 
\usepackage[T1]{fontenc}    
\usepackage{hyperref}       
\usepackage{url}            
\usepackage{booktabs}       
\usepackage{amsfonts}       
\usepackage{nicefrac}       
\usepackage{microtype}      

\usepackage[pdftex]{color,graphicx}
\usepackage{amsmath}
\usepackage{amssymb}
\newcommand{\bs}[1] {\boldsymbol{#1}}
\DeclareMathOperator*{\argmin}{arg\,min}

\title{A Forward Model at Purkinje Cell Synapses Facilitates Cerebellar Anticipatory Control}

\author{
  Ivan Herreros-Alonso \\
  SPECS lab\\
  Universitat Pompeu Fabra\\
  Barcelona, Spain \\
  \texttt{ivan.herreros@upf.edu} \\
  \And 
  Xerxes D. Arsiwalla \\
  SPECS lab\\
  Universitat Pompeu Fabra\\
  Barcelona, Spain \\
  \And 
  Paul F.M.J. Verschure \\
  SPECS, UPF \\
  Catalan Institution of Research\\ and Advanced Studies (ICREA) \\
  Barcelona, Spain  
}

\begin{document}

\maketitle

\begin{abstract}
How does our motor system solve the problem of anticipatory control in spite of a wide spectrum of response dynamics from different musculo-skeletal systems, transport delays as well as response latencies throughout the central nervous system? To a great extent, our highly-skilled motor responses are a result of a reactive feedback system, originating in the brain-stem and spinal cord, combined with a feed-forward anticipatory system, that is adaptively fine-tuned by sensory experience and originates in the cerebellum. Based on that interaction we design the counterfactual predictive control (CFPC) architecture, an anticipatory adaptive motor control scheme in which a feed-forward module, based on the cerebellum, steers an error feedback controller with \emph{counterfactual} error signals. Those are signals that trigger reactions as actual errors would, but that do not code for any current or forthcoming errors. In order to determine the optimal learning strategy, we derive a novel learning rule for the feed-forward module that involves an eligibility trace and operates at the synaptic level. In particular, our eligibility trace provides a mechanism beyond co-incidence detection in that it convolves a history of prior synaptic inputs with error signals. In the context of cerebellar physiology, this solution implies that Purkinje cell synapses should generate eligibility traces using a forward model of the system being controlled. From an engineering perspective, CFPC provides a general-purpose anticipatory control architecture equipped with a learning rule that exploits the full dynamics of the closed-loop system.

\end{abstract}

\section{Introduction}

Learning and anticipation are central features of cerebellar computation and function \citep{Bastian2006}: the cerebellum learns from experience and is able to anticipate events, thereby complementing a reactive feedback control by an anticipatory feed-forward one \citep{hofstoetter2002cerebellum, Herreros2013a}.
This interpretation is based on a series of anticipatory motor behaviors that originate in the cerebellum. For instance, anticipation is a crucial component of acquired behavior in eye-blink conditioning \citep{Gormezano1983}, a trial by trial learning protocol where an initially neutral stimulus such as a tone or a light (the conditioning stimulus, CS) is followed, after a fixed delay, by a noxious one, such as an air puff to the eye (the unconditioned stimulus, US). During early trials, a protective unconditioned response (UR), a blink, occurs reflexively in a feedback manner following the US. After training though, a well-timed anticipatory blink (the conditioned response, CR) precedes the US. Thus, learning results in the (partial) transference from an initial feedback action to an anticipatory (or predictive) feed-forward one. Similar responses occur during anticipatory postural adjustments, which are postural changes that precede voluntary motor movements, such as raising an arm while standing \citep{Massion1992}. The goal of these anticipatory adjustments is to counteract the postural and equilibrium disturbances that voluntary movements introduce. These behaviors can be seen as feedback reactions to events that after learning have been transferred to feed-forward actions anticipating the predicted events.

Anticipatory feed-forward control can yield high performance gains over feedback control whenever the feedback loop exhibits transmission (or transport) delays \citep{jordan1996computational}. However, even if a plant has negligible transmission delays, it may still have sizable inertial latencies. For example, if we apply a force to a visco-elastic plant, its peak velocity will be achieved after a certain delay; i.e. the velocity itself will lag the force. An efficient way to counteract this lag will be to apply forces anticipating changes in the desired velocity. That is, anticipation can be beneficial even when one can act instantaneously on the plant. Given that, here we address two questions: what is the optimal strategy to learn anticipatory actions in a cerebellar-based architecture? and how could it be implemented in the cerebellum?

To answer that we design the counterfactual predictive control (CFPC) scheme, a cerebellar-based adaptive-anticipatory control architecture that learns to anticipate performance errors from experience. The CFPC scheme is motivated from neuro-anatomy and physiology of eye-blink conditioning. It includes a reactive controller, which is an output-error feedback controller that models brain stem reflexes actuating on eyelid muscles, and a feed-forward adaptive component that models the cerebellum and learns to associate its inputs with the error signals driving the reactive controller. With CFPC we propose a generic scheme in which a feed-forward module enhances the performance of a reactive error feedback controller steering it with signals that facilitate anticipation, namely, with \emph{counterfactual errors}. However, within CFPC, even if these counterfactual errors that enable predictive control are learned based on past errors in behavior, they do not reflect any current or forthcoming error in the ongoing behavior.


In addition to eye-blink conditioning and postural adjustments, the interaction between reactive and cerebellar-dependent acquired anticipatory behavior has also been studied in paradigms such as  visually-guided smooth pursuit eye movements \citep{Lisberger1987}. All these paradigms can be abstracted as tasks in which the same predictive stimuli and disturbance or reference signal are repeatedly experienced. In accordance to that, we operate our control scheme in trial-by-trial (batch) mode. With that, we derive a learning rule for anticipatory control that modifies the well-known least-mean-squares/Widrow-Hoff rule with an eligibility trace. More specifically, our model predicts that to facilitate learning, parallel fibers to Purkinje cell synapses implement a forward model that generates an eligibility trace. Finally, to stress that CFPC is not specific to eye-blink conditioning, we demonstrate its application with a smooth pursuit task.

\section{Methods}
\subsection{Cerebellar Model}
\begin{figure}[h]
  \centering
  \includegraphics[width=0.4\textwidth]{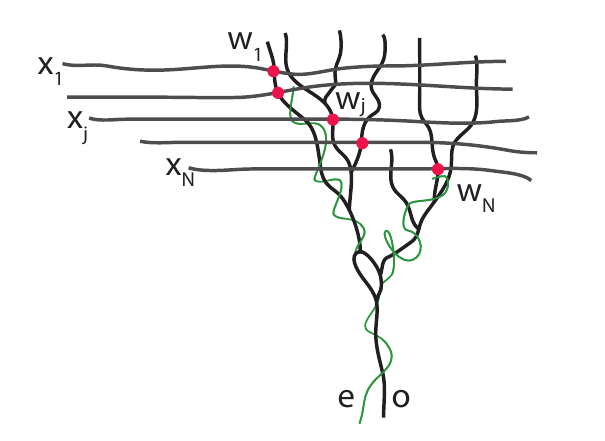}
  \caption{Anatomical scheme of a Cerebellar Purkinje cell. The $x_j$ denote parallel fiber inputs to Purkinje synapses (in red) with weights $w_j$. $o$ denotes the output of the Purkinje cell. The error signal $e$, through the climbing fibers (in green), modulates synaptic weights. }
\label{Purkinje}
\end{figure}
We follow the simplifying approach of modeling the cerebellum as a linear adaptive filter, while focusing on computations at the level of the Purkinje cells, which are the main output cells of the cerebellar cortex \citep{Fujita1982,Dean2010}. Over the mossy fibers, the cerebellum receives a wide range of inputs. Those inputs reach Purkinke cells  via parallel fibers (Fig.~\ref{Purkinje}), that cross dendritic trees of Purkinje cells in a ratio of up to  $1.5 \times 10^6$ parallel fiber synapses per cell \citep{Eccles67}. We denote the signal carried by a particular fiber as $x_j$, $j \in [1,G]$, with $G$ equal to the total number of inputs fibers. These inputs from the mossy/parallel fiber pathway carry contextual information (interoceptive or exteroceptive) that allows the Purkinje cell to generate a functional output. We refer to these inputs as \emph{cortical bases}, indicating that they are localized at the cerebellar cortex and that they provide a repertoire of states and inputs that the cerebellum combines to generate its output $o$. As we will develop a discrete time analysis of the system, we use $n$ to indicate time (or time-step). The output of the cerebellum at any time point $n$ results from a weighted sum of those cortical bases. $w_j$ indicates the weight or synaptic efficacy associated with the fiber $j$. Thus, we have $\bs{x}[n] = \left[ x_1[n], \dots , x_G[n] \right]^\intercal$ and $\bs{w}[n]=\left[ w_1[n], \dots , w_G[n] \right]^\intercal$ (where the transpose, $^\intercal$, indicates that $\bs{x}[n]$ and $\bs{w}[n]$ are column vectors) containing the set of inputs and synaptic weights at time $n$, respectively, which determine the output of the cerebellum according to 
\begin{equation}
o[n]=\bs{x}[n]^\intercal\bs{w}[n]
\end{equation}
The adaptive feed-forward control of the cerebellum stems from updating the weights according to a rule of the form 
\begin{equation}
\label{defLearningRule}
\Delta w_j[n+1]=f(x_j[n], \dots, x_j[1], e[n],\Theta)
\end{equation}
where $\Theta$ denotes global parameters of the learning rule; $x_j[n], \dots, x_j[1]$, the history of its pre-synaptic inputs of synapse $j$; and $e[n]$, an error signal that is the same for all synapses, corresponding to the difference between the desired, $r$, and the actual output, $y$, of the controlled plant. Note that in drawing an analogy with the eye-blink conditioning paradigm, we use the simplifying convention of considering the noxious stimulus (the air-puff) as a reference, $r$, that indicates that the eyelids should close; the closure of the eyelid as the output of the plant, $y$; and the sensory response to the noxious stimulus as an error, $e$, that encodes the difference between the desired, $r$, and the actual eyelid closures, $y$. Given this, we advance a new learning rule, $f$, that achieves optimal performance in the context of eye-blink conditioning and other cerebellar learning paradigms.

\subsection{Cerebellar Control Architecture}
\begin{figure}[h]
  \centering
  \begin{minipage}[h]{0.45\linewidth} 
\centering
\includegraphics[width=0.99\linewidth]{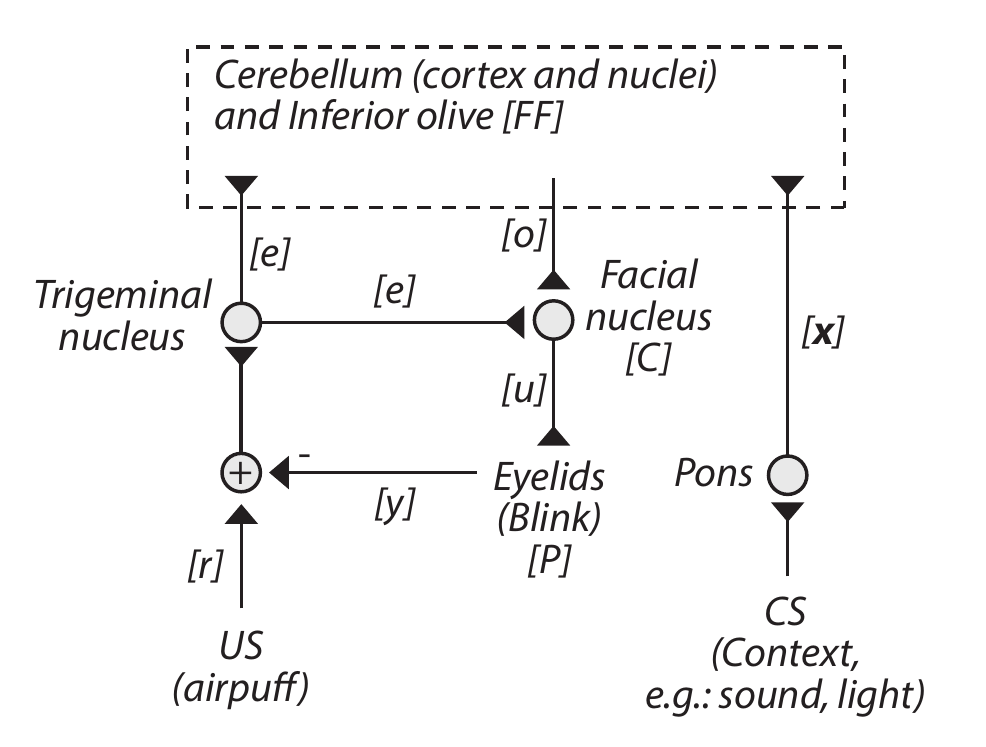}
\end{minipage}%
\begin{minipage}[h]{0.45\linewidth} 
\centering
\includegraphics[width=0.99\linewidth]{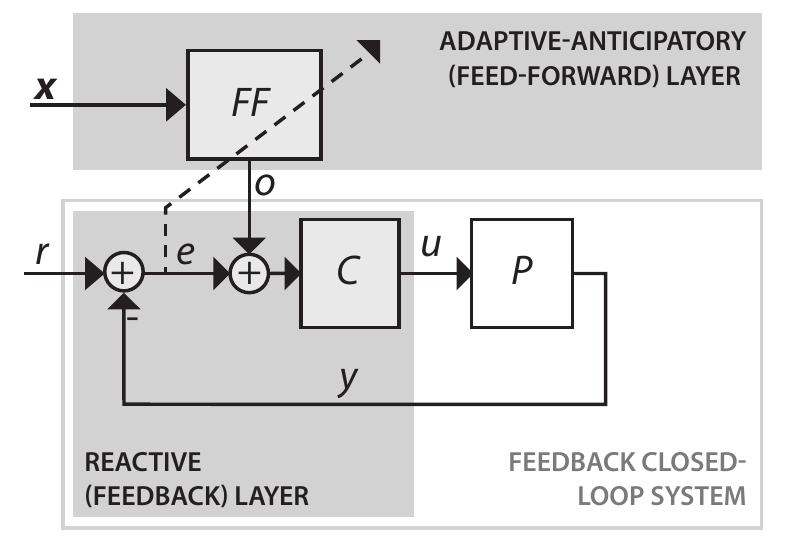} 
\end{minipage}
\caption{Neuroanatomy of eye-blink conditioning and the CFPC architecture. \emph{Left}: Mapping of signals to anatomical structures in eye-blink conditioning \citep{DeZeeuw2005}; regular arrows indicate external inputs and outputs, arrows with inverted heads indicate neural pathways. \emph{Right}: CFPC architecture. Note that the feedback controller, $C$, and the feed-forward module, $FF$, belong to the control architecture, while the plant, $P$, denotes an object controlled. Other abbreviations: $r$, reference signal; $y$, plant's output; $e$, output error; $\mathbf{x}$, basis signals; $o$, feed-forward signal; and $u$, motor command.}
\label{anatomy}
\end{figure}
We embed the adaptive filter cerebellar module in a layered control architecture, namely the CFPC architecture, based on the interaction between brain stem motor nuclei driving motor reflexes and the cerebellum, such as the one established between the cerebellar microcircuit responsible for conditioned responses and the brain stem reflex circuitry that produces unconditioned eye-blinks \citep{hesslow2002functional} (Fig.~\ref{anatomy} \emph{left}). Note that in our interpretation of this anatomy we assume that cerebellar output, $o$, feeds the lower reflex controller (Fig.~\ref{anatomy} \emph{right}). Put in control theory terms, within the CFPC scheme an adaptive feed-forward layer supplements a negative feedback controller steering it with feed-forward signals. 

Our architecture uses a single-input single-output negative-feedback controller. The controller receives as input the output error $e=r-y$. For the derivation of the learning algorithm, we assume that both plant and controller are linear and time-invariant (LTI) systems. 
Importantly, the feedback controller and the plant form a reactive closed-loop system, that mathematically can be seen as a system that maps the reference, $r$, into the plant's output, $y$. 
A feed-forward layer that contains the above-mentioned cerebellar model provides the negative feedback controller with an additional input signal, $o$. We refer to $o$ as a \emph{counter-factual} error signal, since although it \emph{mechanistically} drives the negative feedback controller analogously to an error signal it is not an \emph{actual} error. The counterfactual error is generated by the feed-forward module that receives an output error, $e$, as its teaching signal. Notably, from the point of view of the reactive layer closed-loop system, $o$ can also be interpreted as a signal that offsets $r$. In other words, even if $r$ remains the reference  that sets the target of behavior, $r+o$ functions as the \emph{effective} reference that drives the closed-loop system.

\section{Results}
\subsection{Derivation of the gradient descent update rule for the cerebellar control architecture}
\label{Derivation}
We apply the CFPC architecture defined in the previous section to a task that consists in following a finite reference signal $\mathbf{r} \in \mathbb{R}^N$ that is repeated trial-by-trial. To analyze this system, we use the discrete time formalism and assume that all components are linear time-invariant (LTI). Given this, both reactive controller and plant can be lumped together into a closed-loop dynamical system, that can be described with the dynamics $\mathbf{A}$, input $\mathbf{B}$, measurement $\mathbf{C}$ and feed-through $\mathbf{D}$ matrices. In general, these matrices describe how the state of a dynamical system autonomously evolves with time, $\mathbf{A}$; how inputs affect system states, $\mathbf{B}$; how states are mapped into outputs, $\mathbf{C}$; and how inputs instantaneously affect the system's output $\mathbf{D}$ \citep{Astrom2012}.  
As we consider a reference of a finite length $N$, we can construct the $N$-by-$N$ transfer matrix $\mathcal{T}$ as follows \citep{Boyd2008}
\[\mathcal{T} = \left[\begin{array}{cccccc}
\bs{D}                &      0         &  0     & \hdots &   0     \\
\bs{C} \bs{B}         &      \bs{D}    &  0     & \hdots &   0     \\
\bs{C} \bs{A}  \bs{B} & \bs{C} \bs{B}  & \bs{D} & \hdots &   0     \\
\vdots                & \vdots         & \vdots & \ddots & \vdots \\
\bs{C} \bs{A}^{N-2} \bs{B}  &  \bs{C} \bs{A}^{N-3} \bs{B}  &  \bs{C} \bs{A}^{N-4} \bs{B} & \hdots & \bs{D}  
\end{array}\right]\]
With this transfer matrix we can map any given reference $\mathbf{r}$ into an output $\mathbf{y}_r$ using $\mathbf{y}_r=\mathcal{T} \mathbf{r}$, obtaining what would have been the complete output trajectory of the plant on an entirely feedback-driven trial. Note that the first column of $\mathcal{T}$ contains the impulse response curve of the closed-loop system, while the rest of the columns are obtained shifting that impulse response down. Therefore, we can build the transfer matrix $\mathcal{T}$ either in a model-based manner, deriving the state-space characterization of the closed-loop system, or in measurement-based manner, measuring the impulse response curve. Additionally, note that $(\mathbf{I}-\mathcal{T})\mathbf{r}$ yields the error of the feedback control in following the reference, a signal which we denote with $\mathbf{e}_0$. 

Let $\mathbf{o} \in \mathbb{R}^N$ be the entire feed-forward signal for a given trial. Given commutativity, we can consider that from the point of view of the closed-loop system $o$ is added directly to the reference $\mathbf{r}$, (Fig.~\ref{anatomy} \emph{right}). In that case, we can use $\mathbf{y}=\mathcal{T}(\mathbf{r}+\mathbf{o})$ to obtain the output of the closed-loop system when it is driven by both the reference and the feed-forward signal. 
The feed-forward module only outputs linear combinations of a set of bases. Let $\mathbf{X} \in \mathbb{R}^{N \times G}$ be a matrix with the content of the $G$ bases during all the $N$ time steps of a trial. The feed-forward signal becomes $\mathbf{o}=\mathbf{Xw}$, where $\mathbf{w} \in \mathbb{R}^G$ contains the mixing weights. Hence, the output of the plant given a particular $\mathbf{w}$ becomes $\mathbf{y}=\mathcal{T}(\mathbf{r}+\mathbf{Xw})$.

We implement learning as the process of adjusting the weights $\mathbf{w}$ of the feed-forward module in a trial-by-trial manner. At each trial the same reference signal, $\mathbf{r}$, and bases, $\mathbf{X}$, are repeated. Through learning we want to converge to the optimal weight vector $\mathbf{w}^*$ defined as
\begin{equation}
\mathbf{w}^* = \argmin_w c(\mathbf{w}) = \argmin_w \frac{1}{2} \mathbf{e}^\intercal \mathbf{e}= \argmin_w \frac{1}{2} (\mathbf{r}-\mathcal{T}(\mathbf{r}+\mathbf{Xw}))^\intercal(\mathbf{r}-\mathcal{T}(\mathbf{r}+\mathbf{Xw}))
\end{equation}
where $c$ indicates the objective function to minimize, namely the $L_2$ norm or sum of squared errors. With the substitution $\tilde{\mathbf{X}}=\mathcal{T}\mathbf{X}$ and using  $\mathbf{e}_0 = (\mathbf{I}-\mathcal{T})\mathbf{r}$, the minimization problem can be cast as a \emph{canonical} linear least-squares problem: 
\begin{equation}
\mathbf{w}^* = \argmin_w \frac{1}{2}
 (\mathbf{e}_0-\tilde{\mathbf{X}}\mathbf{w})^\intercal(\mathbf{e}_0-\tilde{\mathbf{X}}\mathbf{w})
\end{equation}
One the one hand, this allows to directly find the least squares solution for $\mathbf{w}^*$, that is, $\mathbf{w}^*=\tilde{\mathbf{X}}^\dagger \mathbf{e}_0$, where $\dagger$ denotes the Moore-Penrose pseudo-inverse. On the other hand, and more interestingly, with $\mathbf{w}[k]$ being the weights at trial $k$ and having $\mathbf{e}[k] = \mathbf{e}_0-\tilde{\mathbf{X}}\mathbf{w}[k]$, we can obtain the gradient of the error function at trial $k$ with relation to $\bs{w}$ as follows:
\[
\nabla_w c = -\tilde{\mathbf{X}}^\intercal \mathbf{e}[k] = -\mathbf{X}^\intercal \mathcal{T}^\intercal \ \mathbf{e}[k]
\]
Thus, setting $\eta$ as a properly scaled learning rate (the only global parameter $\Theta$ of the rule), we can derive the following gradient descent strategy for the update of the weights between trials:
\begin{equation}
\label{eqUpdate}
\mathbf{w}[k+1] = \mathbf{w}[k] + \eta \mathbf{X}^\intercal \mathcal{T}^\intercal \mathbf{e}[k]    
\end{equation}
This solves for the learning rule $f$ in eq.~\ref{defLearningRule}. Note that $f$ is consistent with both the cerebellar anatomy (Fig.~\ref{anatomy}\emph{left}) and the control architecture (Fig.~\ref{anatomy}\emph{right}) in that the feed-forward module/cerebellum only requires two signals to update its weights/synaptic efficacies: the basis inputs, $\mathbf{X}$, and error signal, $\mathbf{e}$.

\subsection{$\mathcal{T}^\intercal$ facilitates a synaptic eligibility trace}
The standard least mean squares (LMS) rule (also known as  Widrow-Hoff or decorrelation learning rule) can be represented in its batch version as $\mathbf{w}[k+1] = \mathbf{w}[k] + \eta \mathbf{X}^\intercal \mathbf{e}[k]$. Hence, the only difference between the batch LMS rule and the one we have derived is the insertion of the matrix factor $\mathcal{T}^\intercal$. Now we will show how this factor acts as a filter that computes an eligibility trace at each weight/synapse. 
Note that the update of a single weight, according Eq.~\ref{eqUpdate} becomes 
\begin{equation}
w_j[k+1] = w_j[k] + \eta \mathbf{x}_j^\intercal \mathcal{T}^\intercal \mathbf{e}[k]
\end{equation} 
where $\mathbf{x}_j$ contains the sequence of values of the cortical basis $j$ during the entire trial. This can be rewritten as 
\begin{equation}
\label{errorAsAvector}
w_j[k+1] = w_j[k] + \eta \mathbf{h}_j^\intercal \mathbf{e}[k]
\end{equation}
with $\mathbf{h}_j \equiv \mathcal{T} \mathbf{x}_j$. The above inner product can be expressed as a sum of scalar products \begin{equation}
\label{errorAsAscalar}
w_j[k+1] = w_j[k] + \eta \sum_{n=1}^N \mathbf{h}_j[n] \mathbf{e}[k,n]
\end{equation}
where $n$ indexes the within trial time-step. Note that $\mathbf{e}[k]$ in Eq.~\ref{errorAsAvector} refers to the whole error signal at trial $k$ whereas $\mathbf{e}[k,n]$ in Eq.~\ref{errorAsAscalar} refers to the error value in the $n$-th time-step of the trial $k$. It is now clear that each $\mathbf{h}_j[n]$ weighs how much an error arriving at time $n$ should modify the weight $w_j$, which is precisely the role of an eligibility trace. Note that since $\mathcal{T}$ contains in its columns/rows shifted repetitions of the impulse response curve of the closed-loop system, the eligibility trace codes at any time $n$, the convolution of the sequence of previous inputs with the impulse-response curve of the reactive layer closed-loop. Indeed, in each synapse, the eligibility trace is generated by a forward model of the closed-loop system that is exclusively driven by the basis signal. 

Consequently, our main result is that by deriving a gradient descent algorithm for the CFPC cerebellar control architecture we have obtained an exact definition of the suitable eligibility trace. That definition guarantees that the set of weights/synaptic efficacies are updated in a locally optimal manner in the weights' space.

\subsection{On-line gradient descent algorithm}

The trial-by-trial formulation above allowed for a straightforward derivation of the (batch) gradient descent algorithm. As it lumped together all computations occurring in a same trial, it accounted for time within the trial implicitly rather than explicitly: one-dimensional time-signals were mapped onto points in a high-dimensional space. However, after having established the gradient descent algorithm, we can implement the same rule in an on-line manner, dropping the repetitiveness assumption inherent to trial-by-trial learning and performing all computations locally in time.
Each weight/synapse must have a process associated to it that outputs the eligibility trace. That process passes the incoming (unweighted) basis signal through a (forward) model of the closed-loop as follows: 
\[\begin{array}{rcl}
\bs{s}_j[n+1] & = & \bs{A} \bs{s}_j[n] + \bs{B} x_j[n] \\
h_j[n] & = & \bs{C} \bs{s}_j[n] + \bs{D} x_j[n]
\end{array}
\]
where matrices $\bs{A}$, $\bs{B}$, $\bs{C}$ and $\bs{D}$ refer to the closed-loop system (they are the same matrices that we used to define the transfer matrix $\mathcal{T}$), and $\bs{s}_j[n]$ is the state vector of the forward model of the synapse $j$ at time-step $n$. In practice, each ``synaptic'' forward model computes what would have been the effect of having driven the closed-loop system with each basis signal alone. Given the superposition principle, the outcome of that computation can also be interpreted as saying that $h_j[n]$ indicates what would have been the displacement over the current output of the plant, $y[n]$, achieved feeding the closed-loop system with the basis signal $x_j$.
The process of weight update is completed as follows: 
\begin{equation}
w_j[n+1] = w_j[n] + \eta h_j[n] e[n]
\end{equation}
At each time step $n$, the error signal $e[n]$ is multiplied by the current value of the eligibility trace $h_j[n]$, scaled by the learning rate $\eta$, and subtracted to the current weight $w_j[n]$. Therefore whereas the contribution of each basis to the output of the adaptive filter depends only on its current value and weight, the change in weight depends on the current and past values passed through a forward model of the closed-loop dynamics.

\subsection{Simulation of a visually-guided smooth pursuit task}
We demonstrate the CFPC approach in an example of a visual smooth pursuit task in which the eyes have to track a target moving on a screen. Even though the simulation does not capture all the complexity of a smooth pursuit task, it illustrates our anticipatory control strategy. We model the plant (eye and ocular muscles) with a two-dimensional linear filter that maps motor commands into angular positions. Our model is an extension of the model in \citep{Porrill2007}, even though in that work the plant was considered in the context of the vestibulo-ocular reflex. In particular, we use a chain of two leaky integrators: a slow integrator with a relaxation constant of 100 ms drives the eyes back to the rest position; the second integrator, with a fast time constant of 3 ms ensures that the change in position does not occur instantaneously. To this basic plant, we add a reactive control layer modeled as a proportional-integral (PI) error-feedback controller, with proportional gain $k_p$ and integral gain $k_i$. The control loop includes a 50 ms delay in the error feedback, to account for both the actuation and the sensing latency. We choose gains such that reactive tracking lags the target by approximately 100 ms. This gives $k_p=20$ and $k_i=100$. To complete the anticipatory and adaptive control architecture, the closed-loop system is supplemented by the feed-forward module. 

\begin{figure}[h]
\label{Results1}
  \centering
  \begin{minipage}[h]{0.4\linewidth} 
\centering
\includegraphics[width=0.99\linewidth]{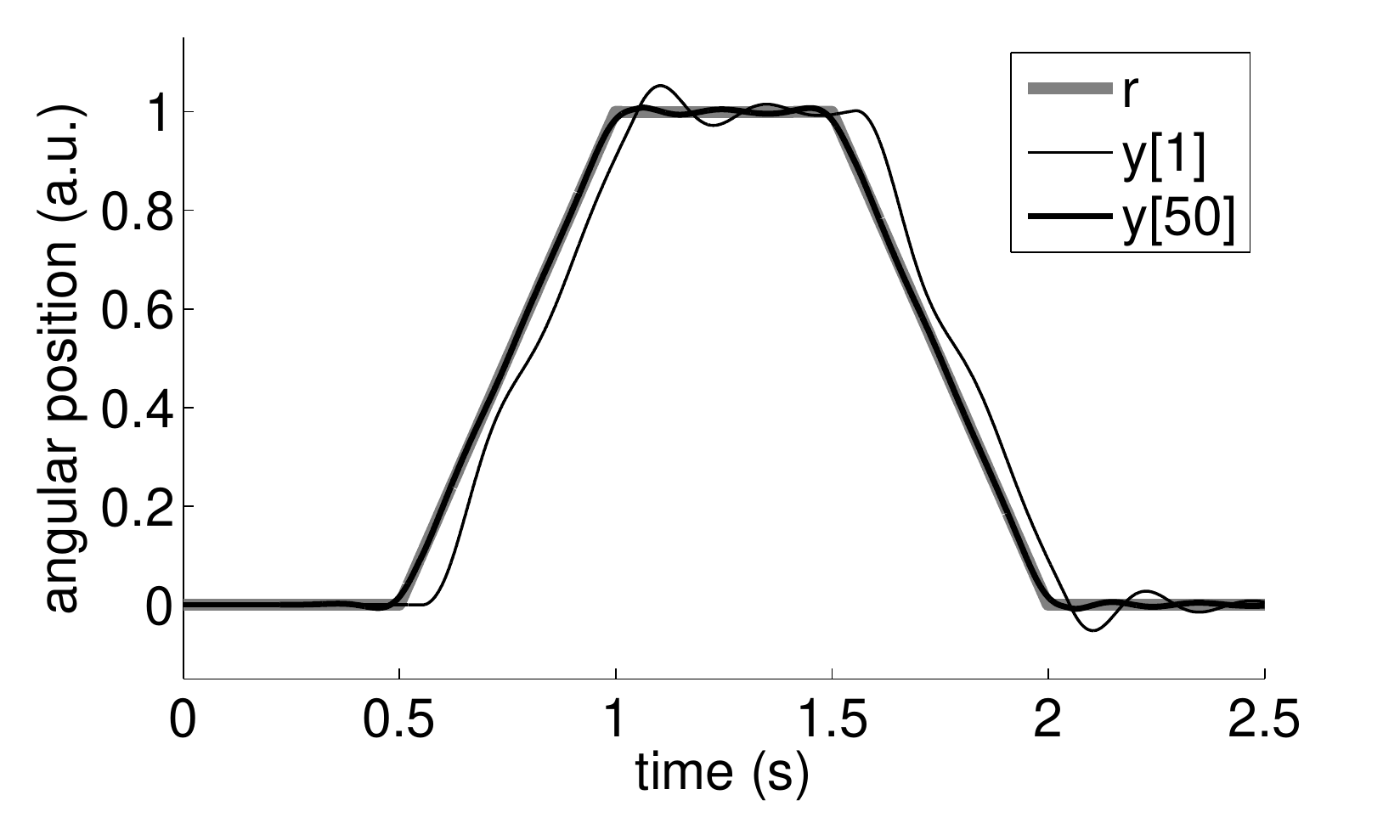}
\end{minipage}%
\begin{minipage}[h]{0.4\linewidth} 
\centering
\includegraphics[width=0.99\linewidth]{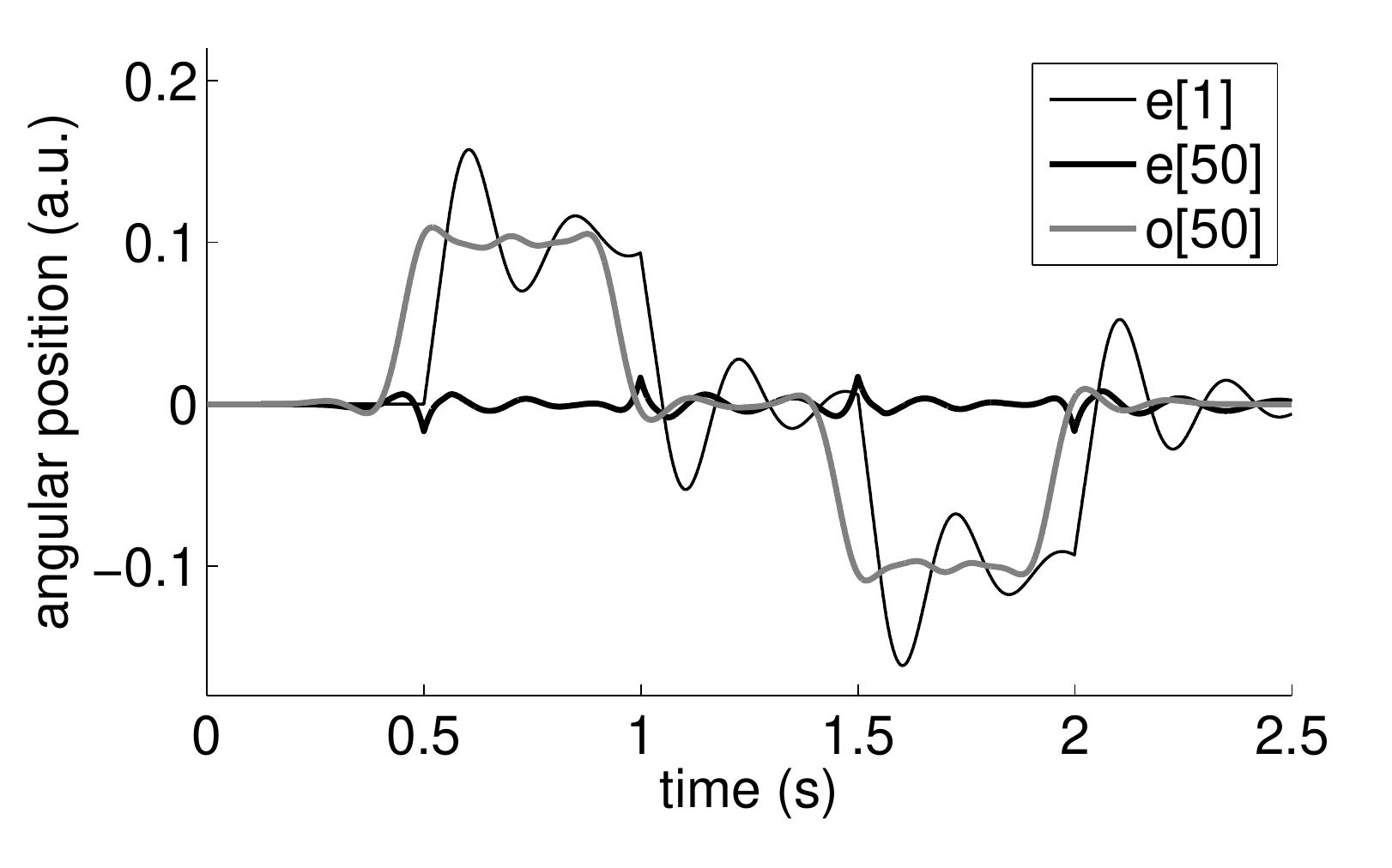} 
\end{minipage}
\caption{Behavior of the system. Left: Reference ($\mathbf{r}$) and output of the system before ($\mathbf{y}[1]$) and after learning ($\mathbf{y}[50]$). Right: Error before $\mathbf{e}[1]$ and after learning $\mathbf{e}[50]$ and output acquired by cerebellar/feed-forward component ($\mathbf{o}[50]$)}
\label{fig1}
\end{figure}

The architecture implementing the forward model-based gradient descent algorithm is applied to a task structured in trials of $2.5$ sec duration. Within each trial, a target remains still at the center of the visual scene for a duration 0.5 sec, next it moves rightwards for 0.5 sec with constant velocity, remains still for 0.5 sec and repeats the sequence of movements in reverse, returning to the center. The cerebellar component receives 20 Gaussian basis signals ($\mathbf{X}$) whose receptive fields are defined in the temporal domain, relative to trial onset, with a width (standard-deviation) of $50$ ms and spaced by $100$ ms. The whole system is simulated using a $1$ ms time-step. To construct the matrix $\mathcal{T}$ we computed closed-loop system impulse response.

\begin{figure}[h]
\label{Results2}
  \centering
\begin{minipage}[h]{0.5\linewidth} 
\centering
\includegraphics[width=0.99\linewidth]{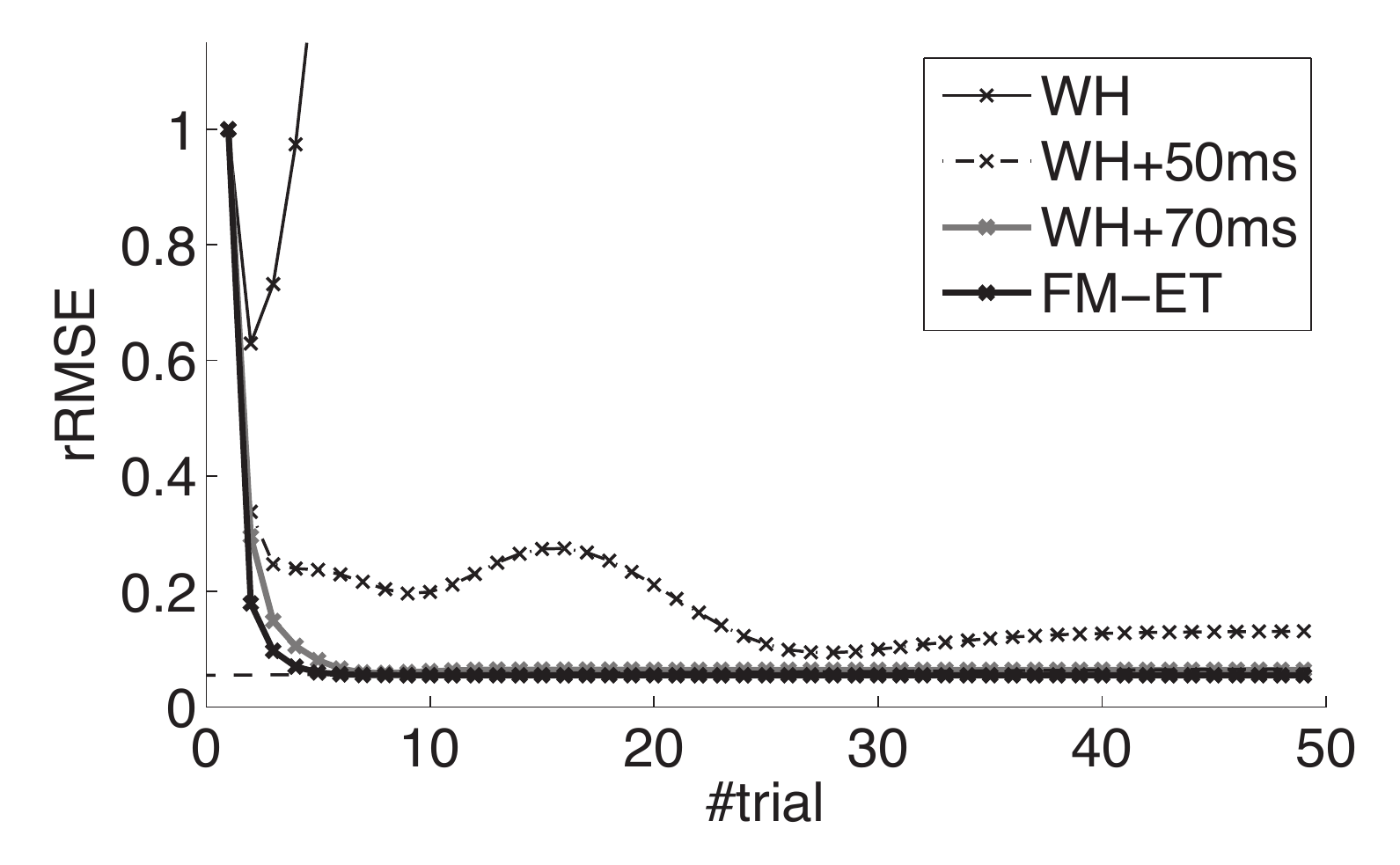} 
\end{minipage}
\caption{Performance achieved with different learning rules. Representative learning curves of the forward model-based eligibility trace gradient descent (FM-ET), the simple Widrow-Hoff (WH) and the Widrow-Hoff algorithm with a delta-eligibility trace matched to error feedback delay (WH+50 ms) or with an eligibility trace exceeding that delay by 20 ms (WH+70 ms). Error is quantified as the relative root mean-squared error (rRMSE), scaled proportionally to the error in the first trial. Error of the optimal solution, obtained with $\bs{w}^*=(\mathcal{T}\bs{X})^\dagger \bs{e}_0$, is indicated with a dashed line.}
\label{fig2}
\end{figure}

At the first trial, before any learning, the output of the plant lags the reference signal by approximately $100$ ms converging to the position only when the target remains still for about $300$ ms (Fig. \ref{fig1} \emph{left}). As a result of learning, the plant's behavior  shifts from a reactive to an anticipatory mode, being able to track the reference without any delay. Indeed, the error that is sizable during the target displacement before learning,  almost completely disappears by the 50$^{th}$ trial (Fig. \ref{fig1} \emph{right}). That cancellation results from  learning the weights that generate a feed-forward predictive signal that leads the changes in the reference signal (onsets and offsets of target movements) by approximately $100$ ms (Fig. \ref{fig1} \emph{right}). Indeed, convergence of the algorithm is remarkably fast and by trial $7$ it has almost converged to the optimal solution (Fig. \ref{fig2}). 

To assess how much our forward-model-based eligibility trace contributes to performance, we test three alternative algorithms. In both cases we employ the same control architecture, changing the plasticity rule such that we either use no eligibility trace, thus implementing the basic Widrow-Hoff learning rule, or use the Widrow-Hoff rule extended with a delta-function eligibility trace that matches the latency of the error feedback (50 ms) or slightly exceeds it (70 ms). Performance with the basic WH model worsens rapidly whereas performance with the WH learning rule using a ``pure delay'' eligibility trace matched to the transport delay improves but not as fast as with the forward-model-based eligibility trace (Fig. \ref{fig2}).  Indeed, in this case, the best strategy for implementing a delayed delta eligibility trace is setting a delay exceeding the transport delay by around 20 ms, thus matching the peak of the impulse response. In that case, the system performs almost as good as with the forward-model eligibility trace (70 ms). This last result implies that, even though the literature usually emphasizes the role of transport delays, eligibility traces also account for response lags due to intrinsic dynamics of the plant. 

To summarize our results, we have shown with a basic simulation of a visual smooth pursuit task that generating the eligibility trace by means of a forward model ensures convergence to the optimal solution and accelerates learning by guaranteeing that it follows a gradient descent. 

\section{Discussion}

In this paper we have introduced a novel formulation of cerebellar anticipatory control, consistent with experimental evidence, in which a forward model has emerged naturally at the level of Purkinje cell synapses.  From a machine learning perspective, we have also provided an optimality argument for the derivation of an eligibility trace, a construct that was often thought of in more heuristic terms as a mechanism to bridge time-delays \citep{barto1983neuronlike,shibata2001biomimetic,mckinstry2006cerebellar}.

The first seminal works of cerebellar computational models emphasized its role as an associative memory \citep{Marr1969,albus1971theory}. Later, the cerebellum was investigates as a device processing correlated time signals\citep{Fujita1982,Kawato1987,Dean2010}. In this latter framework, the use of the computational concept of an eligibility trace emerged as a heuristic construct that allowed to compensate for transmission delays in the circuit\citep{Kettner1997, shibata2001biomimetic, Porrill2007}, which introduced lags in the cross-correlation between signals. Concretely, that was referred to as the problem of \emph{delayed error feedback},  due to which, by the time an error signal reaches a cell, the synapses accountable for that error are no longer the ones currently active, but those that were active at the time when the motor signals that caused the actual error were generated. This view has however neglected the fact that beyond transport delays, response dynamics of physical plants also influence how past pre-synaptic signals could have related to the current output of the plant. Indeed, for a linear plant, the impulse-response function of the plant provides the complete description of how inputs will drive the system, and as such, integrates transmission delays as well as the dynamics of the plant. Recently, 

Even though cerebellar microcircuits have been used as models for building control architectures, e.g., the feedback-error learning model \citep{Kawato1987}, our CFPC is novel in that it links the cerebellum to the input of the feedback controller, ensuring that the computational features of the feedback controller are exploited at all times. Within the domain of adaptive control, there are remarkable similarities at the functional level between CFPC and iterative learning control (ILC) \citep{amann1996iterative}, which is an input design technique for learning optimal control signals in repetitive tasks. The difference between our CFPC and ILC lies in the fact that ILC controllers directly learn a control signal, whereas, the CFPC learns a conterfactual error signal that steers a feedback controller. However the similarity between the two approaches can help for extending CFPC to more complex control tasks.

With our CFPC framework, we have modeled the cerebellar system at a very high level of abstraction: we have not included bio-physical constraints underlying neural computations, obviated known anatomical connections such as the cerebellar nucleo-olivary inhibition \citep{Bengtsson2006, Herreros2013a} and made simplifications such as collapsing cerebellar cortex and nuclei into the same computational unit. On the one hand, such a choice of high-level abstraction may indeed be beneficial for deriving general-purpose machine learning or adaptive control algorithms. On the other hand, it is remarkable that in spite of this abstraction our framework makes fine-grained predictions at the micro-level of biological processes. Namely, that in a cerebellar microcircuit \citep{Apps2005}, the response dynamics of secondary messengers \citep{wang2000coincidence} regulating plasticity of Purkinje cell synapses to parallel fibers must mimic the dynamics of the motor system being controlled by that cerebellar microcircuit. Notably, the logical consequence of this prediction, that different Purkinje cells should display different plasticity rules according to the system that they control, has been validated recording single Purkinje cells in vivo \citep{suvrathan2016timing}.

In conclusion, we find that a normative interpretation of plasticity rules in Purkinje cell synapses emerges from our systems level CFPC computational architecture. That is, in order to generate optimal eligibility traces, synapses must include a forward model of the controlled subsystem. This conclusion, in the broader picture, suggests that synapses are not merely components of multiplicative gains, but rather the loci of complex dynamic computations that are relevant from a functional perspective, both, in terms of optimizing storage capacity \citep{benna2016computational, lahiri2013memory} and fine-tuning learning rules to behavioral requirements.

\subsubsection*{Acknowledgments}

The research leading to these results has received funding from the European Commission’s Horizon 2020 socSMC project (socSMC-641321H2020-FETPROACT-2014) and by the European Research Council’s CDAC project (ERC-2013-ADG 341196).


\small

\bibliography{NIPS}

\end{document}